%% file: main.tex
\documentclass[conference]{IEEEtran}
\IEEEoverridecommandlockouts

\usepackage{cite}
\usepackage{amsmath,amssymb,amsfonts}
\usepackage{algorithmic}
\usepackage{graphicx}
\usepackage{textcomp}
\usepackage{xcolor}
\def\BibTeX{{\rm B\kern-.05em{\sc i\kern-.025em b}\kern-.08em
    T\kern-.1667em\lower.7ex\hbox{E}\kern-.125emX}}

\usepackage{epstopdf}
\usepackage{lipsum}
\usepackage{makecell}
\usepackage{pdfpages}
\usepackage{tikz}
\usetikzlibrary{shapes,arrows}
\usepackage{pgf-pie}
\usepackage{pgfplots}
\usepgfplotslibrary{groupplots}
\usepackage{verbatim}
\pgfplotsset{compat=1.17}
\usepackage{helvet}
\usepackage[eulergreek]{sansmath}
\usepackage{flushend}
\usepackage{caption}
\usepackage{subcaption}
\usepackage{xcolor}
\usepackage{threeparttable}
\usepackage{multirow}
\usepackage[absolute]{textpos}
\usepackage{hyperref}

\begin{document}

\title{EDEA:
  \underline{E}fficient
  \underline{D}ual-\underline{E}ngine
  \underline{A}ccelerator
  for Depthwise Separable Convolution with Direct Data Transfer
	}

	\author{
		\IEEEauthorblockN{
		    Yi Chen, Jie Lou, Malte Wabnitz, Johnson Loh and Tobias Gemmeke
		}
		\IEEEauthorblockA{
 			Chair of Integrated Digital Systems and Circuit Design,
 			RWTH Aachen University, Germany\\
 			Email: \{chen, lou, wabnitz, loh, gemmeke\}@ids.rwth-aachen.de
		}
	}

    \maketitle

	\input{s0_Abstract}
	\input{s1_Introduction}
	\input{s2_Design_Space_Exploration}

	\input{s3_Hardware_Architecture}
	\input{s4_Experimental_Results}
	\input{s5_Conclusion}

	\section*{Acknowledgment}
	This work is partially funded by the German Federal Ministry of Education and Research under grants 03ZU1106CA (NeuroSys) and 16ME0399 (NEUROTEC II), and by the German Federal Ministry for the Environment, Nature Conservation, Nuclear Safety and Consumer Protection under grant 67KI32006A (RESCALE).
	\bibliographystyle{IEEEtran}
	\bibliography{IEEEabrv,latex/reference}

\end{document}

%% file: s0_Abstract.tex
\begin{textblock*}{\textwidth}(1cm,0.25cm) 
\small{© 2024 IEEE. Personal use of this material is permitted. Permission from IEEE must be
obtained for all other uses, in any current or future media, including
reprinting/republishing this material for advertising or promotional purposes, creating new collective works, for resale or redistribution to servers or lists, or reuse of any copyrighted component of this work in other works. This paper was accepted at the 37th International System-on-Chip Conference (SOCC) 2024. DOI: \href{https://doi.org/10.1109/SOCC62300.2024.10737823}{10.1109/SOCC62300.2024.10737823}} \vspace{0.1cm}
\end{textblock*}
\begin{abstract}
Depthwise separable convolution (DSC) has emerged as a crucial technique, especially for resource-constrained devices.
In this paper, we propose a dual-engine for the DSC hardware accelerator, which enables the full utilization of depthwise convolution (DWC) and pointwise convolution (PWC) processing elements (PEs) in all DSC layers. 
To determine the optimal dataflow, data reuse, and configuration of the target architecture, we conduct a design space exploration using MobileNetV1 with the CIFAR10 dataset.
In the architecture, we introduce an additional non-convolutional unit, which merges the dequantization, batch normalization (BN), ReLU, and quantization between DWC and PWC into a simple fixed-point multiplication and addition operation.
This also reduces the intermediate data access between the DWC and PWC, enabling streaming operation and reducing latency.
The proposed DSC dual-engine accelerator is implemented using the 22nm FDSOI technology from GlobalFoundries, occupying an area of 0.58 \(\text{mm}^2\). 
After signoff, it can operate at 1 GHz at TT corner, 
achieving a peak energy efficiency of 13.43 TOPS/W with a throughput of 973.55 GOPS with 8-bit precision.
The average energy efficiency of all DSC layers on MobileNetV1 is 11.13 TOPS/W, demonstrating substantial hardware efficiency improvements for DSC-based applications.
\end{abstract}

\begin{IEEEkeywords}
    Lightweight CNNs, depthwise separable convolution, dual engines, ASIC accelerator
\end{IEEEkeywords}

%% file: s1_Introduction.tex
\section{Introduction}
Lightweight CNNs leverage depthwise separable convolution (DSC) to streamline model architectures. DSC decomposes standard convolution (SC) into depthwise convolution (DWC) and pointwise convolution (PWC) \cite{Jiang_TCAD_2023}, significantly enhancing model efficiency and reducing computational footprint, which is crucial for AI applications on resource-constrained edge devices.

Efficient hardware implementation is crucial to fully leverage the benefits of DSC.
Although unified convolution engines for both DWC and PWC have been proposed \cite{Yan_2021} \cite{Xuan_2022} \cite{Yi_VLSISoC_2023}, achieving full utilization of processing elements (PEs) for both DWC and PWC remains a challenge. 
The differing proportions of DWC and PWC across layers in various architectures introduce a workload imbalance \cite{Li_TCASI_2021}.
Efforts have been made to address this challenge. 
For instance, a separate engine for DWC and PWC was proposed, enhancing efficiency by mitigating imbalances \cite{preprint_2023}. 
However, this approach does not allow for parallel execution of DWC and PWC.
Although separate DWC and PWC engines capable of parallel operation have been proposed \cite{Wu_2019} \cite{Justin_FPL_2020}, these designs were primarily optimized for FPGA platforms. 
However, FPGAs introduce overhead that obfuscates the actual throughput and energy efficiency potential of a specific architecture \cite{Eyeriss_2017}.
In the end, literature on ASIC implementations for energy-efficient DSC accelerators is scarce.

While DSC effectively reduces the number of parameters, DWC operates as a channel-wise convolution and PWC as an element-wise convolution, both exhibiting limitations in data reuse.
Therefore, the elimination of intermediate data transfer between DWC and PWC becomes crucial for further reduction of data movement.
Given the numerous design options for DSC, employing improper strategies may restrict the circuit-level design space, consequently limiting hardware performance. 
Therefore, comprehensive early-stage design space exploration is vital, coupled with identifying appropriate quantization and mapping strategies to enhance hardware performance \cite{Huang_ACM_2023} \cite{Malte_Memories_2023}. The main contributions of this work are as follows:

\begin{itemize}
    \item  We conduct a design space exploration to characterize the workload requirements of MobileNetV1 \cite{Howard_arXiv_2017} on the CIFAR10 dataset, aiming to achieve balanced operation in reconfigurable PEs. We determine optimal processing directions and PE sizes to minimize data movement and maximize PE utilization across different layers.

    \item A separate DSC engine is proposed. It supports parallel operation for DWC and PWC, reducing latency, and eliminating data movements of intermediate results. Additionally, it maintains 100\% PE utilization for all DSC layers and is friendly to scaling PE sizes without impacting performance.
    
    \item The dequantization, BN, ReLU, and quantization units are merged into a simple fixed-point multiplication and addition operation.
    This approach reduces the overall number of operations, minimizes data movement, and facilitates data transfer between DWC and PWC.

    \item A DSC accelerator is implemented in 22nm FDSOI technology. The post-layout results indicate that the dual-engine DSC accelerator can achieve the peak energy efficiency of 13.43 TOPS/W with a throughput of 973.55 GOPS at 0.8V. The average energy efficiency across all DSC layers on MobiltNetV1 is 11.13 TOPS/W, with an average throughput of 981.42 GOPS.

\end{itemize}

%% file: s2_Design_Space_Exploration.tex
\begin{figure*}[h!]
    \centering
    \includegraphics[width=0.9\textwidth]{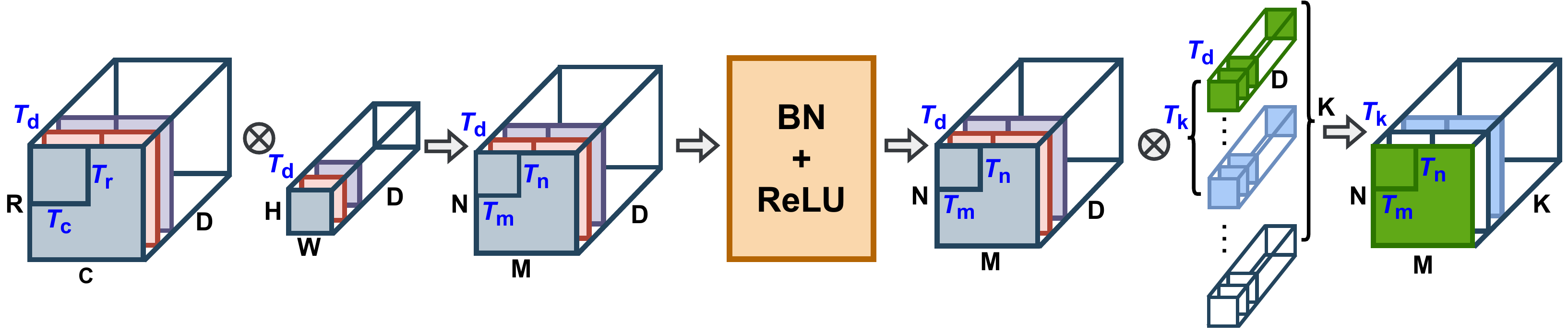}
    \vspace{-0.1cm}
    \caption{Data flow and tile mapping in DWC and PWC.}
    \label{fig:data_flow_direction}
    \vspace{-0.2cm}
\end{figure*}

\section{Design Space Exploration}
The performance of accelerators is significantly impacted by memory footprint, with dataflow and layer tiling playing essential roles in determining computation sequence and optimizing memory utilization. 
However, due to constraints in on-chip resources, it is impractical to store all layer parameters within on-chip memory. 
Hence, identifying optimal dataflow and tile configurations for each layer becomes crucial to maximize accelerator performance \cite{Ryu_ICAD_2024}. 
Fig. \ref{fig:data_flow_direction} depicts a typical DSC layer, with DWC shown on the left and PWC on the right. 
The input feature map (ifmap) size of DWC is $R$$\times$$C$$\times$$D$, with corresponding kernel size $H$$\times$$W$$\times$$D$, resulting in an output feature map (ofmap) size of $N$$\times$$M$$\times$$D$.
The ofmap of the DWC, after batch normalization (BN) and ReLU, servers as the input for the PWC.
It convolves with $K$ kernels of size $1$$\times$$1$$\times$$D$ and yields an output with the size of $N$$\times$$M$$\times$$K$.

To accommodate input data larger than the PE array, the layer should be partitioned into smaller tiles. 
As depicted in Fig. \ref{fig:data_flow_direction}, the process begins with DWC, where the ifmap is partitioned into tiles of size $T_{\rm r}$$\times$$T_{\rm c}$$\times$$T_{\rm d}$, and the kernel is partitioned into tiles of size $H$$\times$$W$$\times$$T_{\rm d}$, resulting in an output dimension of $T_{\rm n}$$\times$$T_{\rm m}$$\times$$T_{\rm d}$. 
The PWC ifmap undergoes convolutions with $T_{\rm k}$ kernels, where each kernel has a tile size of $1$$\times$$1$$\times$$T_{\rm d}$, resulting in a PWC output size $T_{\rm n}$$\times$$T_{\rm m}$$\times$$T_{\rm k}$.

The sequence of data processing will indeed impact data reuse. 
To achieve optimal performance, we investigate various dataflow configurations employing dedicated engines. 
Specifically, five levels of convolution loops are considered.
$\textbf{Loop1}$: MAC within a convolution window size of $T_{r}$$\times$$T_{c}$ for DWC ($T_{n}$$\times$$T_{m}$ for PWC);
$\textbf{Loop2}$: Iterating across the ifmap size of $T_{d}$;
$\textbf{Loop3}$: Scanning the ifmap along the dimensions $R$$\times$$C$ for DWC ($N$$\times$$M$ for PWC);
$\textbf{Loop4}$: Iterating across the ifmap size of $D$;
$\textbf{Loop5}$: Iterating across the ofmap size of $K$.

To determine the optimal design space on the hardware, we conducted an analysis encompassing both tiling size and data processing sequence within the DSC layers of MobileNetV1, utilizing the CIFAR10 dataset.
The loop order is segmented into two distinct sequences, 
$\textbf{La}$: Loop1 $\to$ Loop2 $\to$ Loop3 $\to$ Loop4 $\to$ Loop5;
$\textbf{Lb}$: Loop1 $\to$ Loop2 $\to$ Loop4 $\to$ Loop3 $\to$ Loop5.
Only Loop3 and Loop4 are variable here due to data dependencies, while Loop5 is exclusive to PWC.
To accommodate all layers within the network, especially the later layers such as layers 11 and 12 with an ifmap size of 2, we constrained the tiled size to $T_{\rm n}$=$T_{\rm m}$=1 or 2.
Hence, we categorize our exploration into four distinct groups, encompassing various loop orientations and tiling sizes.
To account for the diverse channel lengths and kernel sizes across different layers, each group is futher subdivided into 6 distinct cases, as outlined in Table \ref{tab:case_method}.
\begin{table}[htbp]
  \centering
  \caption{Seleted tiling sizes.}
  \label{tab:case_method}
  \begin{tabular}{|c|c|c|}
    \hline
    \multirow{2}{*}{Case} & 
    \multirow{2}{*}{$T_{d}$} & 
    \multirow{2}{*}{$T_{k}$}\\
	& & \\
    \hline
    Case1 & 4   &  4  \\
    \hline
    Case2 & 4   &  8  \\
    \hline
    Case3 & 4   &  16  \\
    \hline
    Case4 & 8  &  4  \\
    \hline
    Case5 & 8  & 8  \\
    \hline
    Case6 & 8  & 16  \\
    \hline
  \end{tabular}
  \vspace{-0.3cm}
\end{table}

\begin{figure}[htbp]
    \centering
    \begin{subfigure}[b]{0.48\textwidth}
        \includegraphics[width=1\linewidth]{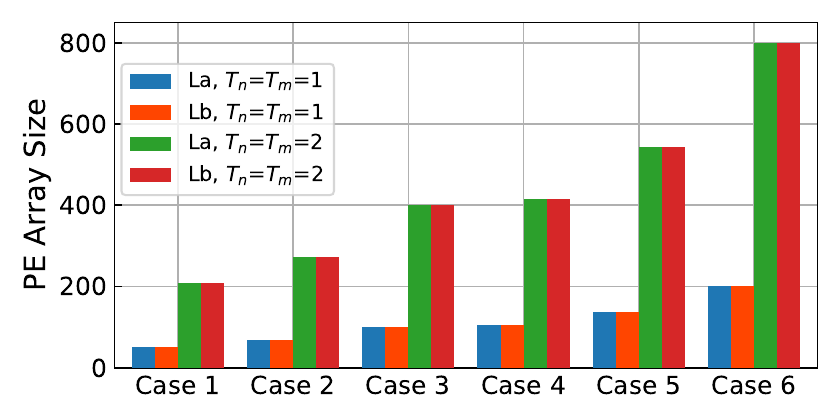}
        \vspace{-0.6cm}
        \caption{}
        \label{fig:PE_size}
    \end{subfigure}
    \begin{subfigure}[b]{0.48\textwidth}
        \includegraphics[width=1\linewidth]{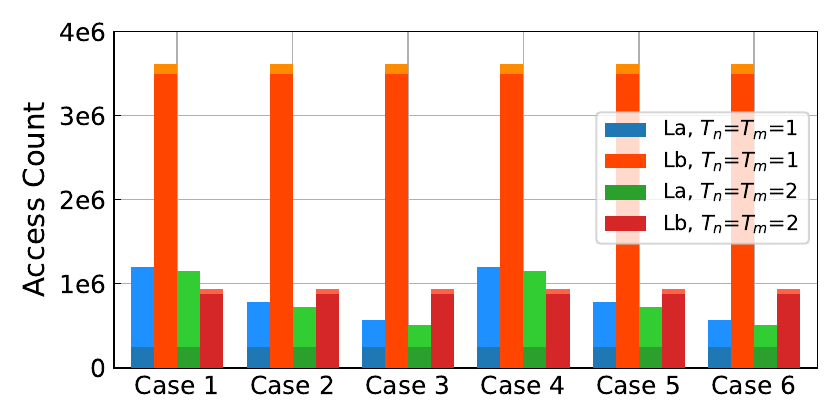}
        \vspace{-0.6cm}
        \caption{}
        \label{fig:access_count}
    \end{subfigure}
    \caption{Design space exploration (a) PE array size (b) Activation and weight access count (upper bar: activation, lower bar: weight).}
    \label{fig:design_space_exploration}
\end{figure}

The memory footprint results of different cases, including all DSC layers of MobileNetV1, are illustrated in Fig. \ref{fig:design_space_exploration}. 
The required PE array size exhibits a linear relationship with the tiling size $T_{\rm n}$, $T_{\rm m}$, $T_{\rm d}$ and $T_{\rm k}$.
The loop order $\textbf{La}$ consistently demonstrates higher activation access count, while $\textbf{Lb}$ consistently exhibits higher weight access count.
For the MobileNetV1 architecture, weight access count significantly outweighs activation access count (Fig. \ref{fig:access_count}). 
Overall, loop order $\textbf{La}$ with $T_{\rm n}$=$T_{\rm m}$=2, in Case6 ($T_{\rm d}$=8, $T_{\rm k}$=16) achieves the lowest access count being our preferred choice for hardware implementation.
The corresponding equations are shown in Table \ref{tab:equations}.

\begin{table}[htbp]
  \centering
  \caption{Equations for loop order La with $T_{\rm n}$=$T_{\rm m}$=2.}
  \label{tab:equations}
  \resizebox{0.49\textwidth}{!}{
  \begin{tabular}{|c|c|c|c|}
    \hline
    Conv & PE Array & Activation Access & Weight Access \\
    \hline
    DWC & $T_{\rm d}$$\times$$H$$\times$$W$$\times$$T_{\rm n}$$\times$$T_{\rm m}$&  
    $T_{\rm r}$$\times$$T_{\rm c}$$\times$$D$$\times$$\frac{N\times M}{T_{\rm n}\times T_{\rm m}}$ & $H$$\times$$W$$\times$$D$ \\
    \hline
    PWC & $T_{\rm d}$$\times$$T_{\rm k}$$\times$$T_{\rm n}$$\times$$T_{\rm m}$ & N$\times$M$\times$D$\times$$\frac{K}{T_{\rm k}}$ & D$\times$K \\
    \hline
  \end{tabular}
 }
  \vspace{-0.3cm}
\end{table}

\begin{figure}[htb]
    \centering
    \includegraphics[width=0.48\textwidth]{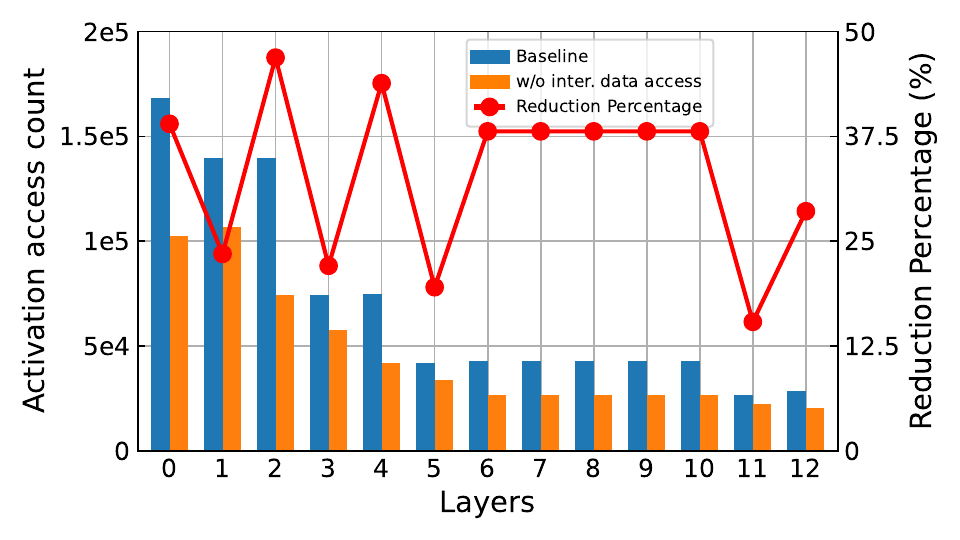}
    \vspace{-0.2cm}
    \caption{Activation access count and reduction percentage}
    \label{fig:proposed_vs_baseline_access}
\end{figure}

To further reduce memory accesses, minimizing intermediate data movement within the DSC layer is crucial.
In Fig. \ref{fig:proposed_vs_baseline_access}, we consider the reduction of memory accesses by eliminating external storage of intermediate activations.
For the baseline, we consider the total activation memory access to include both the input and output memory access of DWC and PWC.
After eliminating the intermediate data access, the total memory access changes to the sum of DWC input and PWC output.
Fig. \ref{fig:proposed_vs_baseline_access} illustrates that eliminating the intermediate data access can achieve an access count reduction ranging from 15.4\% to 46.9\% compared to the baseline across different layers on MobileNetV1, with an total access reduction of 34.7\%.


%% file: s3_Hardware_Architecture.tex
\section{Proposed Hardware Architecture}
\subsection{Overall Architecture}
\begin{figure}[htbp!]
    \centering
    \includegraphics[width=0.42\textwidth]{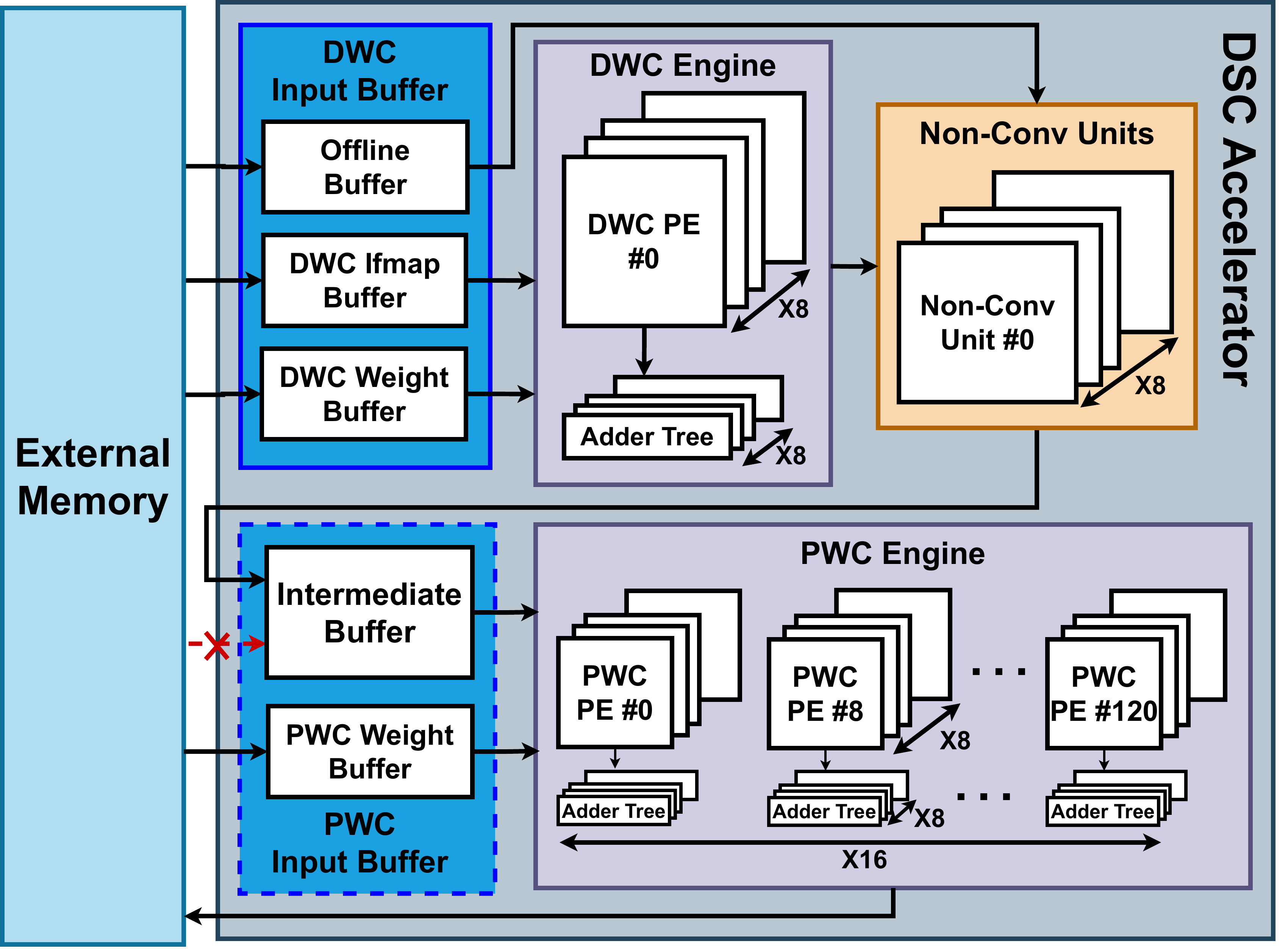}
    \caption{Proposed system architecture.}
    \label{fig:system_architecture}
    \vspace{-0.2cm}
\end{figure}

Figure \ref{fig:system_architecture} illustrates the architecture of the proposed DSC accelerator, which consists of two convolution engines (DWC and PWC), along with a non-convolutional (Non-Conv) unit and two on-chip buffers. 
These two on-chip buffers are dedicated to DWC and PWC, respectively. 
The former consists of the DWC ifmap buffer, DWC weight buffer and an offline buffer dedicated to storing Non-Conv unit parameters.
The latter comprises an intermediate buffer and PWC weight buffer.
To alleviate the necessity of writing DWC activation to external memory and reading PWC activation from external memory, an intermediate buffer is inserted. 
Additionally, to address the BN and ReLU between DWC and PWC, as well as to quantize the PWC input to the required precision, a Non-Conv unit is introduced. 
The unit converts DWC's output into PWC's input by performing a single fixed-point multiplication and addition.

\subsection{Dual Engines}

\begin{figure}[htbp]
    \centering
    \begin{subfigure}[b]{0.44\textwidth}
        \includegraphics[width=1\linewidth]{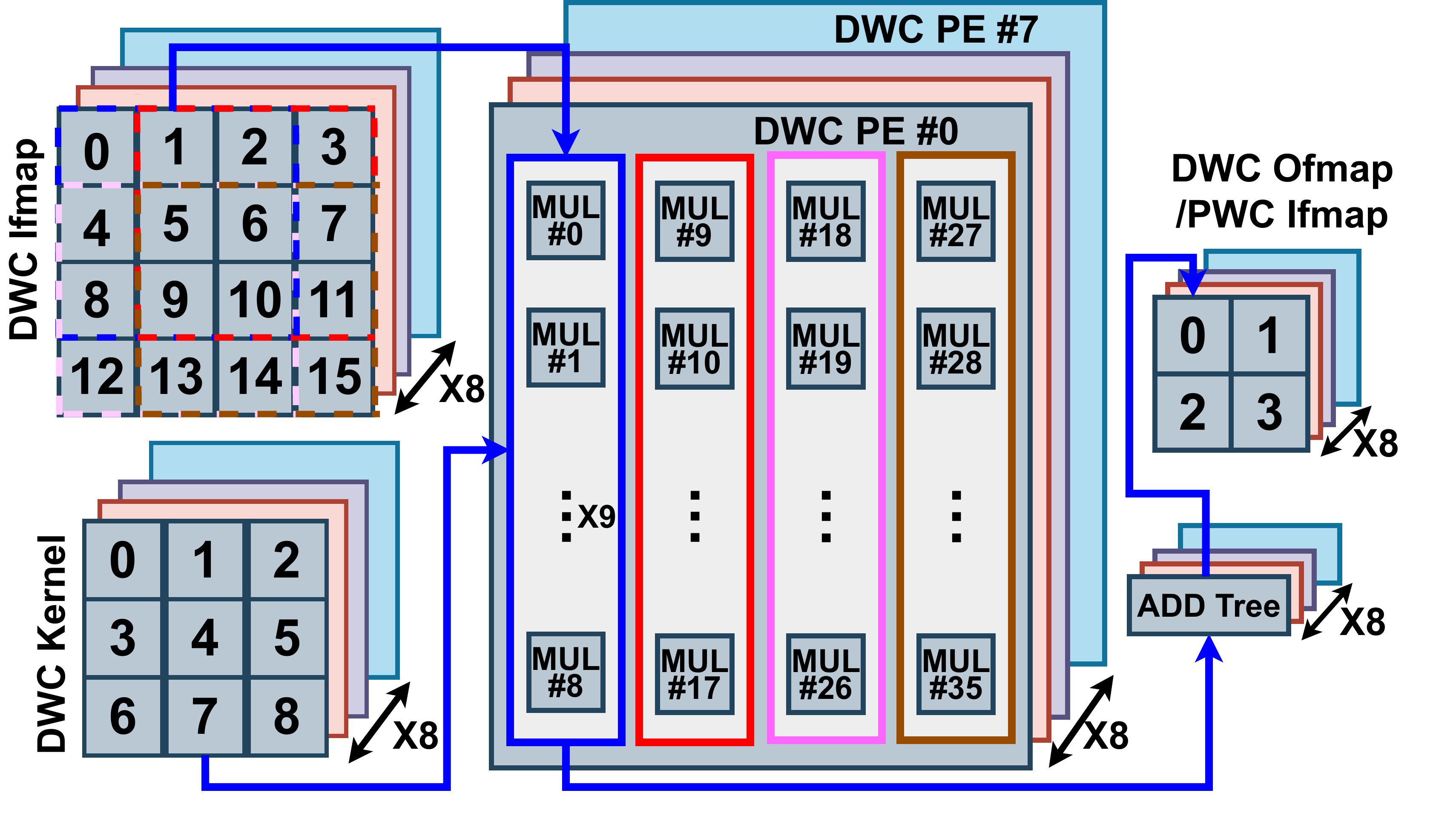}
        \caption{}
        \vspace{0.1cm}
        \label{fig:dwc_engine}
    \end{subfigure}
    \begin{subfigure}[b]{0.43\textwidth}
        \includegraphics[width=1\linewidth]{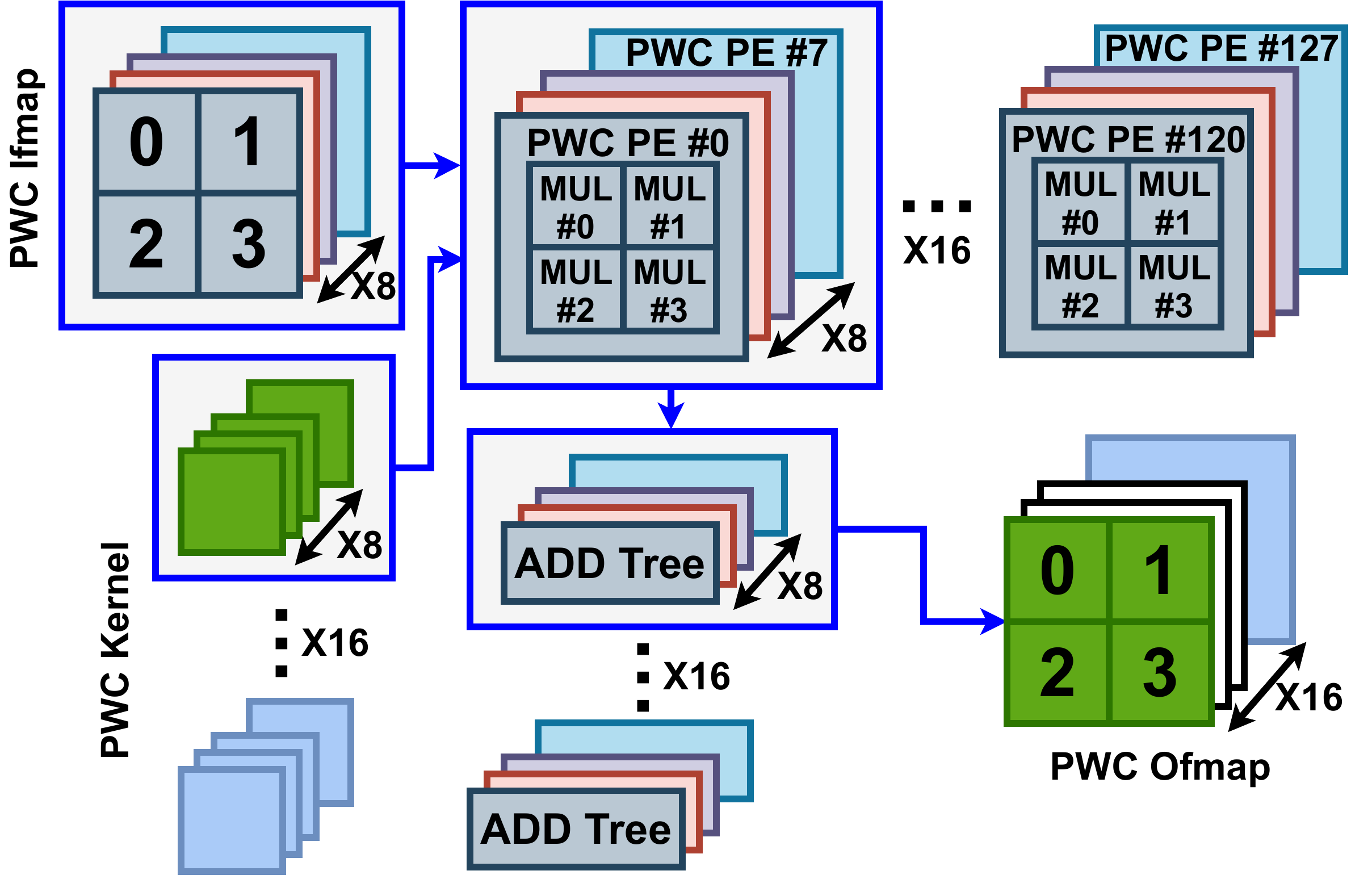}
        \caption{}
        \label{fig:pwc_engine}
    \end{subfigure}
    \caption{DSC dual engines (a) DWC engine (b) PWC engine.}
    \label{fig:dwc_pwc_engine}
\end{figure}

The proposed DWC and PWC engine are depicted in Fig. \ref{fig:dwc_engine} and  Fig. \ref{fig:pwc_engine}.  
The DWC engine consists of a fully parallel PE arrays capable of simultaneously computing 8 channels of ifmap, resulting in a total of 288 MAC operations. 
Each column of PE performs 3$\times$3 MACs using an adder tree and produces the output of DWC, as illustrated in Fig. \ref{fig:dwc_engine}. 
The DWC engine utilizes an ifmap of size 4$\times$4$\times$8 (5$\times$5$\times$8 when stride is 2) and a tiled kernel of size 3$\times$3$\times$8, and generates an ofmap of size 2$\times$2$\times$8.
The DWC ofmap has the same size as the PWC ifmap.
The PWC engine incorporates a total of 512 MAC operations.
It operates on an ifmap with dimensions 2$\times$2$\times$8 and a tiled kernel of size 1$\times$1$\times$8$\times$16, producing an ofmap with dimensions 2$\times$2$\times$16 (Fig. \ref{fig:pwc_engine}).
Notably, this configuration is applicable to all layers of MobileNetV1 while maintaining 100\% PE utilization. 
In addition, PE arrays are friendly to scaling to enhance parallelism without reducing utilization. 
Specifically, in DWC, the number of channels can be scaled, while in PWC, both the number of channels and kernels can be scaled.

\subsection{Non-Convolutional Unit}
\begin{figure}[htbp!]
    \centering
    \includegraphics[width=0.46\textwidth]{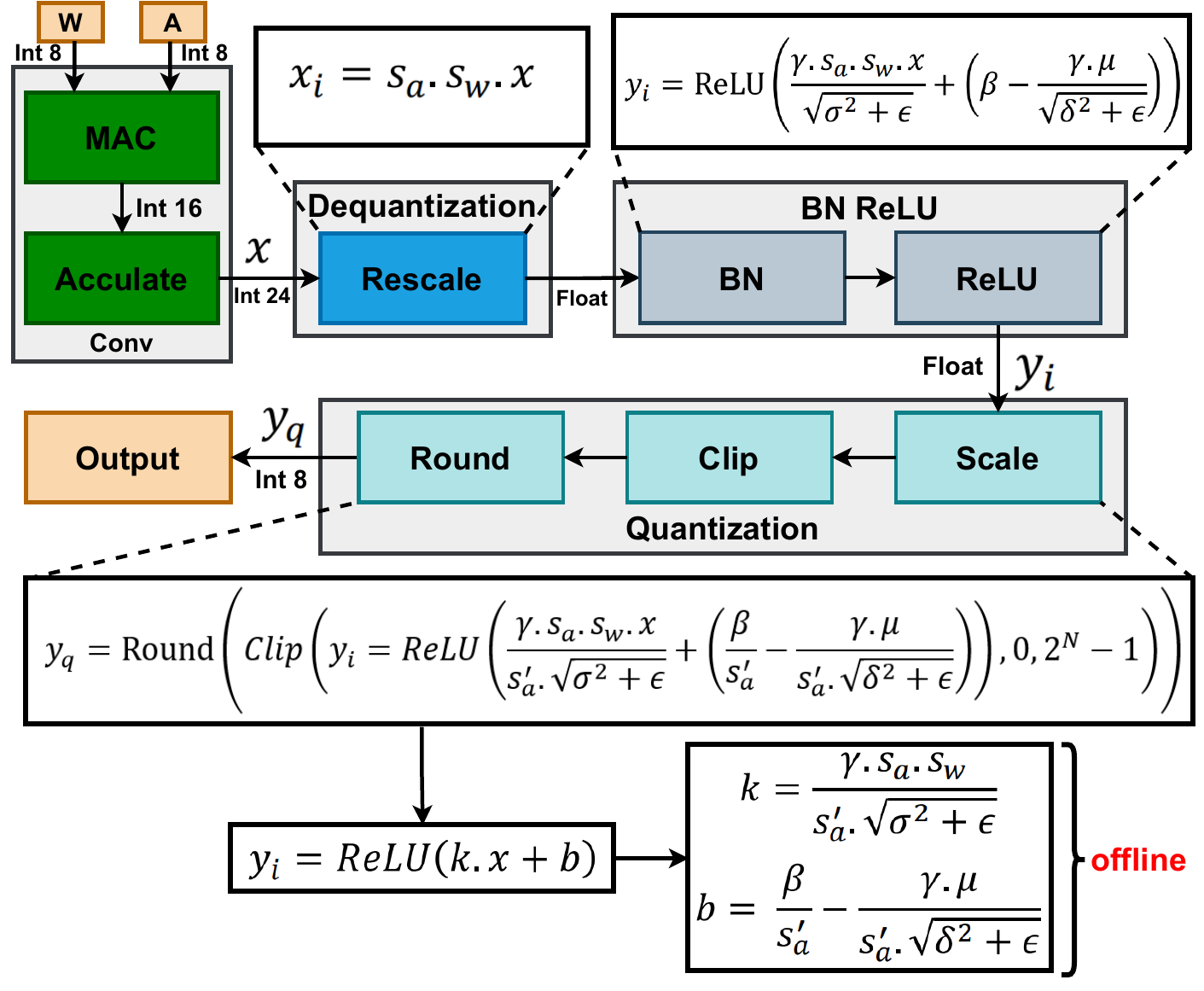}
    \caption{Integration of the Non-Conv unit.}
    \label{fig:non_convolutional_units}
    \vspace{-0.3cm}
\end{figure}

Between DWC and PWC, there are BN and ReLU functions. 
Additionally, to maintains the same word length for PWC activation input as well as DWC activation output, quantization is required.
To facilitate the data transfer from DWC to PWC, a Non-Conv unit is introduced, encompassing dequantization, BN, ReLU and quantization.
In inference, all BN parameters ($\gamma$, $\beta$, $\mu$, $\sigma$, $\epsilon$) and quantization scaling factors ($s_{a}$, $s_{w}$) are fixed and can be pre-computed \cite{Steven_2019} \cite{Zhu_2020}.
As depicted in Fig. \ref{fig:non_convolutional_units}, these parameters and scaling factors can be simplified into a multiplication and addition: $y$=$k$$\times$$x$+$b$, significantly reducing the number of operations.
The introduced Non-Conv unit not only reduces the amount of data needed to be stored in on-chip buffer but also minimizes data movement overhead, facilitating direct data transfer from DWC to PWC.
To cover all possible ranges of the values for $k$ and $b$ without losing precision, we select $k$ and $b$ as 24-bit fixed-point numbers with 8 integer bits and 16 fractional bits.

\subsection{Timing}
\begin{figure}[htbp!]
    \centering
    \includegraphics[width=0.46\textwidth]{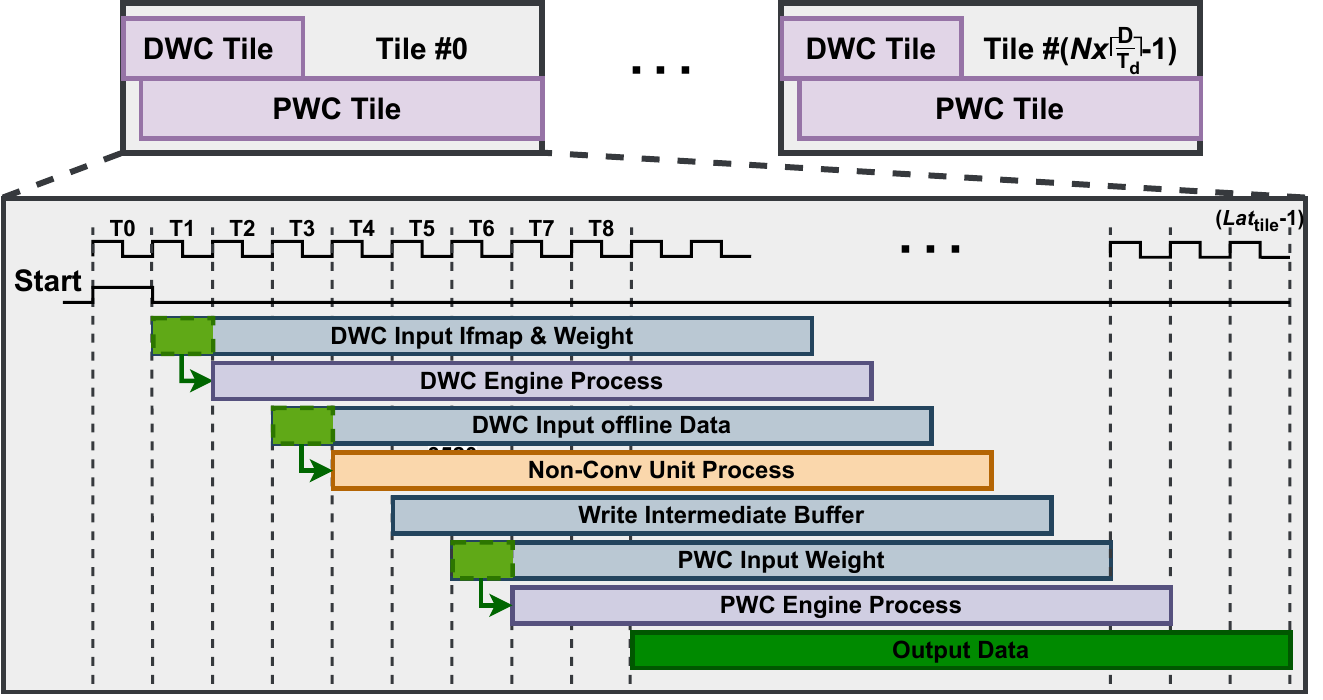}
    \caption{Pipeline timing of separate convolution units.}
    \label{fig:pipeline_timing}
\end{figure}

Figure \ref{fig:pipeline_timing} illustrates the timing of the dual convolution units. 
The DWC and PWC can work in parallel, reducing overall latency and enhancing performance.
As shown in Fig. \ref{fig:pipeline_timing}, the initiation takes 9 clock cycles before generating the first PWC output result.
The tiles latency ($Lat_{\rm tile}$) can be calculated using Eq. \ref{equ:tiles_latency}. 
Due to the limitation posed by the ifmap buffer size, layers with larger dimensions must be partitioned into multiple tiled ifmaps to accommodate the buffer constraints.
The total latency ($Lat_{\rm total}$) of the layer can be calculated using Eq. \ref{equ:total_latency}. where $N$ represents the number of tiled ifmaps.
The DWC PE arrays encounter more idle time due to fewer MAC operations in DWC compared to PWC.
For different layers, the 9 initiation cycles remain constant.
However, due to the change in ifmap/ofmap size, the initiation cycles account for a different contribution.
\begin{equation}
    \label{equ:tiles_latency}
    Lat_{\rm tile} = (9+ \lceil \frac{N}{T_{\rm n}}\rceil \times \lceil \frac{M}{T_{\rm m}}\rceil \times \lceil \frac{K}{T_{\rm k}}\rceil) \times T_{\rm period}
\end{equation}
\begin{equation}
    \label{equ:total_latency}
    Lat_{\rm total} = Lat_{\rm tile} \times N \times \lceil \frac{D}{T_{\rm d}}\rceil
\end{equation}

%% file: s4_Experimental_Results.tex
\section{Experimental Results}
\begin{figure}[hbtp!]
    \centering
    \includegraphics[width=0.24\textwidth]{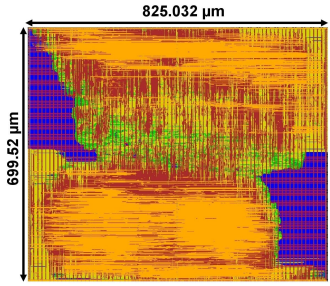}
    \vspace{-0.1cm}
    \caption{Layout view.}
    \label{fig:layout_view}
    \vspace{-0.3cm}
\end{figure}

The proposed DSC accelerator is realized using GlobalFoundries' 22nm FDSOI technology.  
After signoff, it can operate at a clock frequency of 1 GHz at typical (TT) corner.
The accelerator is simulated using QuestaSim, synthesized using Design Compiler, and place and route (P\&R) is done with Innovus. 
The power consumption is conducted using Primetime. 
We trained the MobileNetV1 neural network on the CIFAR10 dataset using the PyTorch framework. 
Afterwards, we quantized the weights and activations to 8 bits using the LSQ \cite{Steven_2019} technique.
Fig. \ref{fig:layout_view} illustrates the layout view of the DSC accelerator, occupying dimensions 825.032\,\textmu m$\times$699.52\,\textmu m. 
The area ratio of PWC to DWC is approximately 1.7X, which closely aligns with the PWC to DWC PE ratio of 1.8X (512 and 288).

\begin{figure}[hbtp!]
    \centering
    \includegraphics[width=0.44\textwidth]{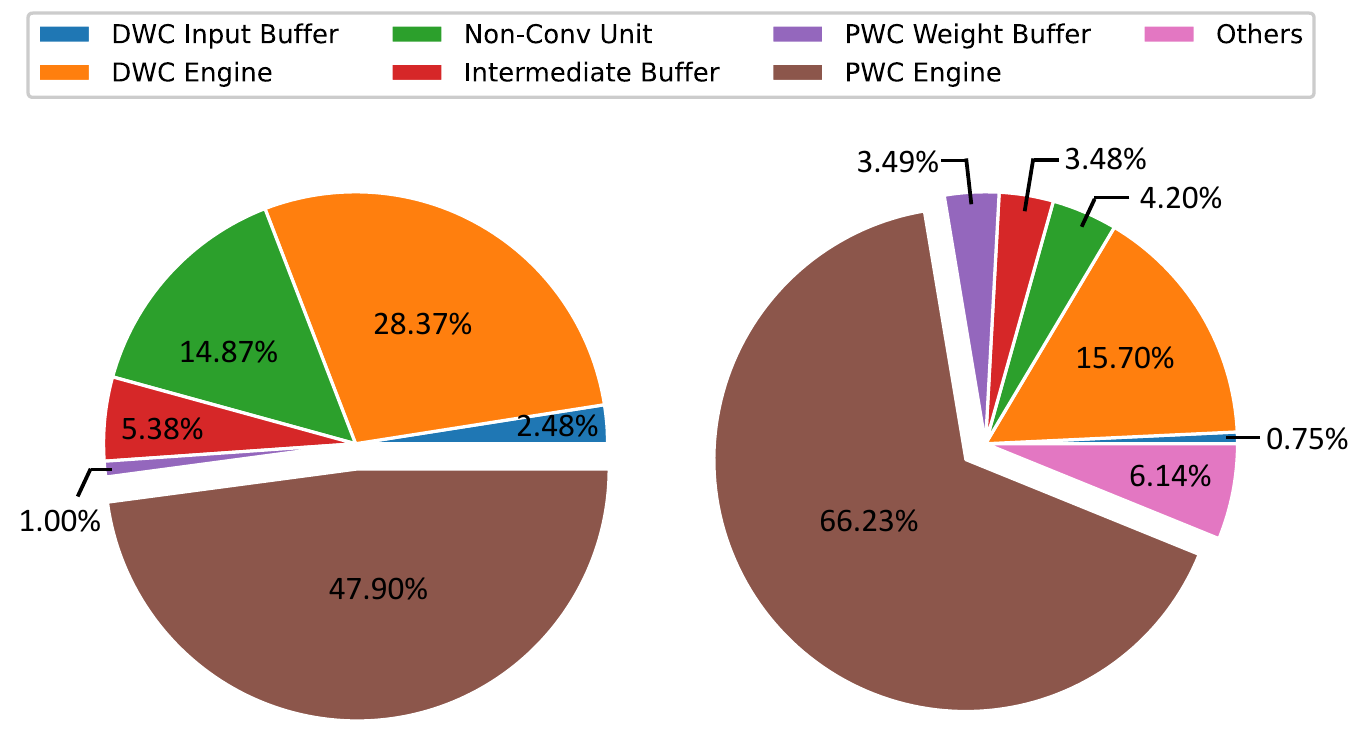}
    \vspace{-0.2cm}
    \caption{Area (left) and power (right) breakdown.}
    \label{fig:area_power_breakdown}
    \vspace{-0.3cm}
\end{figure}

Figure \ref{fig:area_power_breakdown} illustrates the area and power breakdown of the proposed accelerator.
In terms of area, the DWC PE engine occupies 28.37$\%$ of the total area, while the PWC PE engine occupies 47.90$\%$.
Additionally, the Non-Conv unit accounts for 14.87$\%$ of the total area.
Concerning power consumption, the PWC engine accounts for 66.23$\%$ of the total power, followed by the DWC engine at 15.70$\%$. 
The others part in the power breakdown represents the power consumed by the clock tree.

\subsection{Latency and Power}
\begin{figure}[hbtp!]
    \centering
    \includegraphics[width=0.46\textwidth]{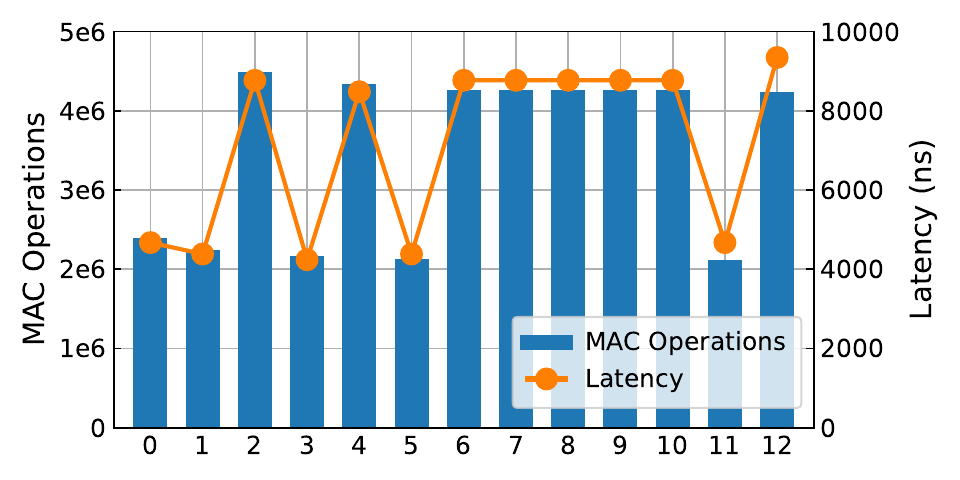}
    \vspace{-0.2cm}
    \caption{MAC operations and latency on different layers.}
    \label{fig:mac_and_total_latency}
    \vspace{-0.3cm}
\end{figure}

\begin{figure}[hbtp!]
    \centering
    \includegraphics[width=0.46\textwidth]{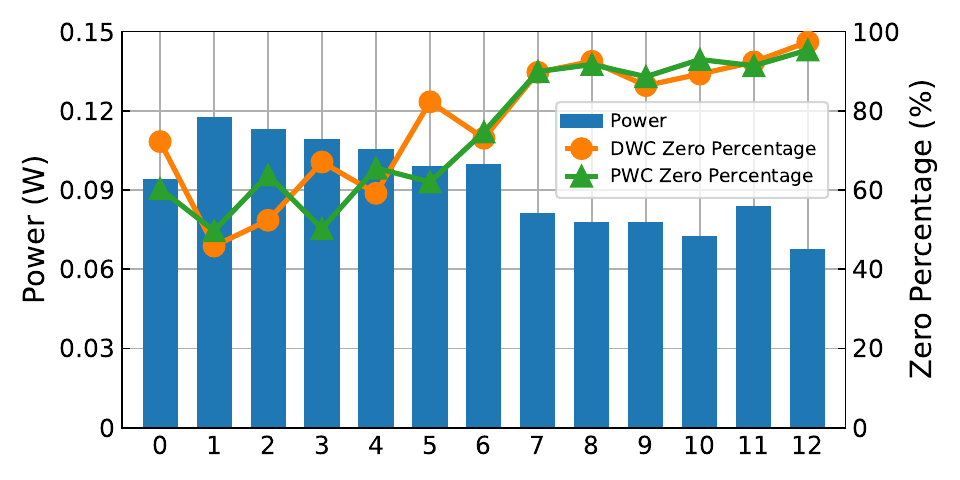}
    \vspace{-0.2cm}
    \caption{Power and zero percentage on different layers.}
    \label{fig:power_and_zero_percentage}
    \vspace{-0.3cm}
\end{figure}

Figure \ref{fig:mac_and_total_latency} illustrates the layerwise MAC operations and total latency. 
There is a strong correlation between the number of MAC operations and the total latency. 
Specifically, layers 1, 3, 5 and 11 exhibit a reduced number of MAC operations due to the stride of 2.
The ifmap/ofmap sizes for later layers are smaller.
According to Eq. \ref{equ:tiles_latency}, the initiation stage, which takes 9 cycles, accounts for a larger contribution, resulting in a slightly increased latency.
Fig. \ref{fig:power_and_zero_percentage} illustrates the power and activation zero percentage of different layers. 
The power reduces as the zero percentage increases.
Layer1 exhibits the highest power consumption of 117.7 mW. 
In contrast, layer12 demonstrates the lowest power consumption of 67.7 mW, while the zero percentage for DWC and PWC are 97.4\% and 95.3\%, respectively. 
\vspace{-0.2cm}

\subsection{Energy Efficiency and Throughput}

\begin{figure}[hbtp!]
    \centering
    \includegraphics[width=0.46\textwidth]{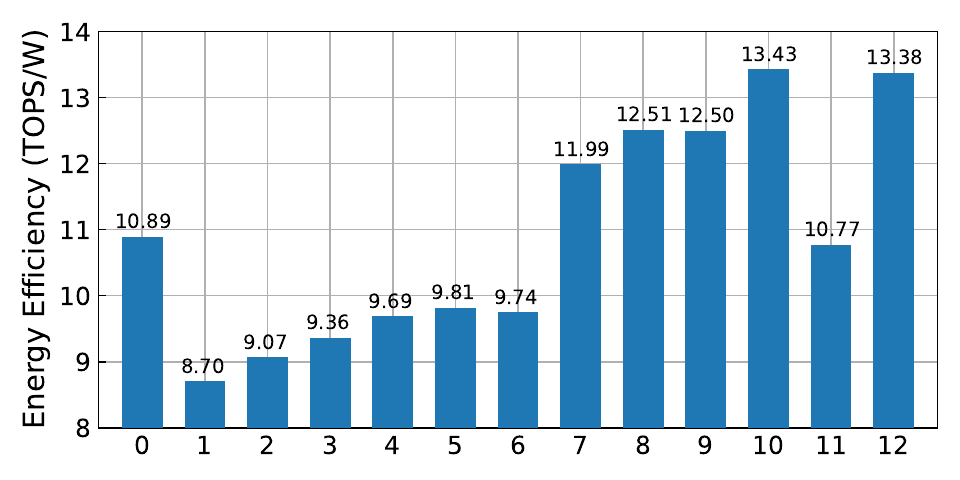}
    \vspace{-0.2cm}
    \caption{Energy efficiency on different layers.}
    \label{fig:layerwise_energy_efficiency}
    \vspace{-0.3cm}
\end{figure}

\begin{figure}[hbtp!]
    \centering
    \includegraphics[width=0.46\textwidth]{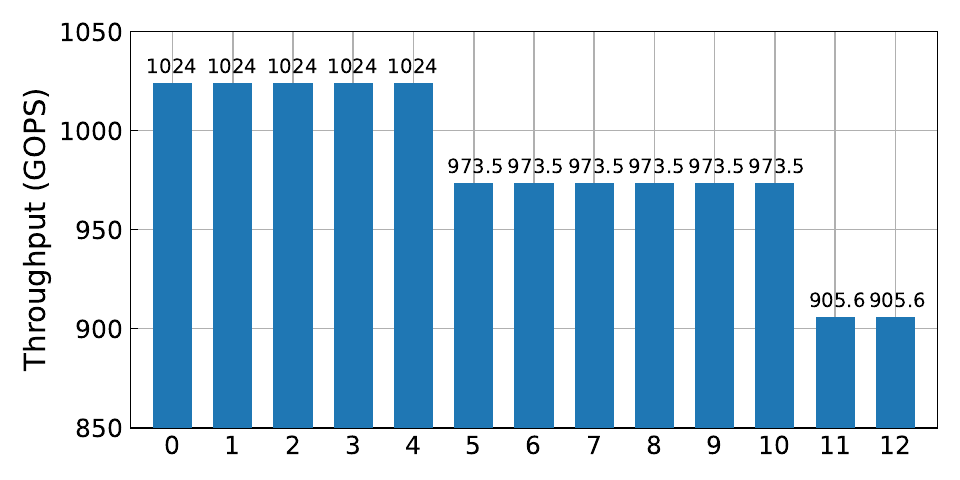}
    \vspace{-0.2cm}
    \caption{Throughput on different layers.}
    \label{fig:layerwise_throughput}
    \vspace{-0.3cm}
\end{figure}

Figure \ref{fig:layerwise_energy_efficiency} shows the energy efficiency across different layers. 
Layer10 demonstrates the highest energy efficiency of 13.43 TOPS/W, attributed to the high zero percentage (cf. Fig. \ref{fig:power_and_zero_percentage}). 
Conversely, layer1 has the lowest energy efficiency of 8.7 TOPS/W. 
The average energy efficiency of all layers is 11.13 TOPS/W.
The throughput across different layers is shown in Fig. \ref{fig:layerwise_throughput}. 
Layers 0 to 4 achieve the highest throughput of 1024 GOPS. 
As the number of layers increases, throughput slightly reduces due to the initiation cycles.
The lowest throughput in layers 11 and 12 is 905.6 GOPS.

\subsection{Comparison with previous works}

 \begin{table*}[h!]
	\centering
	\begin{threeparttable}[b]
	\caption{Comparison with state-of-the-art works.}
	\label{tab:Comp_w_SotA}
	\begin{tabular}{|c||c|c|c|c|c|c|} 
	    \hline
		& ISVLSI'19 \cite{Chen_ISVLSI_2019} &
		TCCE-TW'21 \cite{Hsiao_TCCE-TW_2021} &  
		TCASI'24 \cite{Jung_TCASI_2024} &
        \multicolumn{2}{c|}{VLSI-SoC'23 \cite{Yi_VLSISoC_2023}} & 
		This Work \\
		\hline
		\hline
		Technology & 65\,nm  & 40\,nm  & 28\,nm & \multicolumn{2}{c|}{22\,nm} & 22\,nm  \\ 		
		\hline
		Precision & 8 bit & 16 bit & 8 bit & \multicolumn{2}{c|}{8 bit} & 8 bit \\
		\hline
		Voltage (V) & 1.08  & 0.9 & 0.9  & \multicolumn{2}{c|}{0.8} & 0.8 \\ 
		\hline
		PE Count & 256 & 128 & 288  & \multicolumn{2}{c|}{72} & 800 \\
		\hline
 		Benchmark & MobileNetV1  & MobileNetV1  & DTN\tnote{\textdagger} & \multicolumn{2}{c|}{MobileNetV1} &  MobileNetV1\\  
 		\hline
 		Conv Type & DWC+PWC  & DWC+PWC & SC+DSC & DWC & PWC & DWC+PWC \\
 		\hline
 		Power  (mW) & 55.4  & 112.5  & 43.6 & 25.6  & 29.16 & 72.5  \\
        \hline
        Frequency(MHz) & 100 & 200  & 200 & 1000  & 1000 & 1000  \\
        \hline
        Area (${\rm mm}^2$) & 3.24  & 2.168  & 1.485 & 0.25  & 0.25  & 0.58  \\
		\hline
		Throughput (GOPS) & 51.2  & 38.8 (155.2\tnote{\textdaggerdbl} ) & 215.6 & 129.8 & 115.38 & 973.55 \\
		\hline
		Energy Efficiency & \multirow{2}{*}{0.92}   & \multirow{2}{*}{0.34 (1.36\tnote{\textdaggerdbl} )}  & \multirow{2}{*}{4.94}  & \multirow{2}{*}{5.07} & \multirow{2}{*}{3.96} & \multirow{2}{*}{13.43}    \\
		(TOPS/W) & & & & & &  \\
		\hline
		Area Efficiency & \multirow{2}{*}{15.8} & \multirow{2}{*}{17.9 (71.6\tnote{\textdaggerdbl} )}  & \multirow{2}{*}{145.28} & \multirow{2}{*}{519.2} & \multirow{2}{*}{461.52} & \multirow{2}{*}{1678.53} \\
		 (GOPS/${\rm mm}^2$) & & & & & & \\
		\hline		
        Normalized Energy & \multirow{2}{*}{7.73} & \multirow{2}{*}{1.08 (4.32\tnote{\textdaggerdbl} )}  & \multirow{2}{*}{9.9} & \multirow{2}{*}{5.07} & \multirow{2}{*}{3.96} & \multirow{2}{*}{13.43}   \\
        Efficiency (TOPS/W) & & & & & & \\
		\hline
        Normalized Area & \multirow{2}{*}{266.86}  & \multirow{2}{*}{72.53 (290.12\tnote{\textdaggerdbl} )}  & \multirow{2}{*}{255} & \multirow{2}{*}{519.2} & \multirow{2}{*}{461.52} & \multirow{2}{*}{1678.53} \\
        Efficiency (GOPS/${\rm mm}^2$) & & & & & & \\
		\hline		
	\end{tabular}
	\begin{tablenotes}
	   \item [\textdagger] Depthfused trilateral network  \quad
	   \textdaggerdbl \hspace*{0.01cm} Normalized to 8 bits using $(\frac{\rm Precision}{8})^2$
	\end{tablenotes}	
	\end{threeparttable}
	\vspace{-0.3cm}
\end{table*}

Table \ref{tab:Comp_w_SotA} presents the comparison with state-of-the-art (SotA) works. 
To ensure a fair comparison, all selected references are based on post P\&R results and peak performance.
Our work achieves the peak energy efficiency of 13.43 TOPS/W with a throughput of 973.55 GOPS and an area efficiency of 1678.53 GOPS/$\text{mm}^2$.
Before technology scaling, our work surpasses \cite{Chen_ISVLSI_2019}, \cite{Hsiao_TCCE-TW_2021}, \cite{Jung_TCASI_2024}, \cite{Yi_VLSISoC_2023} by 14.6X, 9.87X, 2.72X, 2.65X in energy efficiency, respectively. Post-scaling to 22nm at 0.8V following the methodology in  \cite{Cecilia_Access_2021}, our study maintains its advantage, outperforming them by 1.74X, 3.11X, 1.37X, 2.65X in energy efficiency and by 6.29X, 7.79X, 6.58X, 3.23X in area efficiency, respectively. Our research illustrates superior energy and area efficiency relative to existing works.
Specifically, when compared to \cite{Yi_VLSISoC_2023}, which realized a unified DSC engine using the same 22nm technology, our work not only demonstrates better results but also maintains high utilization and performance for all layers.

%% file: s5_Conclusion.tex
\section{Conclusion}
In this paper, we first conduct design space exploration using the MobileNetV1 architecture on the CIFAR10 dataset to identify the optimal dataflow and tiling size considering all layers.
We then present a DSC accelerator with dual engines for DWC and PWC, supporting streaming operation and achieving 100\% PE utilization.
A Non-Conv unit is introduced between DWC and PWC to facilitate data transfer, reduce data movement, and simplify computation.
Our results demonstrate the peak energy efficiency of 13.43 TOPS/W and the peak throughput of 1024 GOPS, with an average energy efficiency and throughput of 11.13 TOPS/W and 981.42 GOPS, respectively. 
This dataflow is applicable to other datasets, and the accelerator is also suitable for other DSC-based networks.